\documentclass[lettersize,journal]{IEEEtran}
\usepackage{amsmath,amsfonts}
\usepackage{algorithmic}
\usepackage{array}
\usepackage[caption=false,font=normal,labelfont=rm,textfont=rm]{subfig}
\usepackage{textcomp}
\usepackage{multirow}
\usepackage{stfloats}
\usepackage{url}
\usepackage{verbatim}
\usepackage{diagbox}
\usepackage{subfig} 
\usepackage{subfloat}
\usepackage{float}
\usepackage{algorithm, algorithmic}
\usepackage{amssymb}
\usepackage{amsmath,amsfonts,amsthm,bm}
\usepackage[justification=centering]{caption}
\usepackage{cite}
\usepackage{soul}
\usepackage{makecell}
\usepackage{enumitem}
\usepackage[colorlinks,linkcolor=magenta,anchorcolor=magenta,citecolor=magenta,bookmarks=true]{hyperref} 
\usepackage[capitalize]{cleveref}
\usepackage{graphicx}
\renewcommand\arraystretch{1.3}
\def\BibTeX{{\rm B\kern-.05em{\sc i\kern-.025em b}\kern-.08em
    T\kern-.1667em\lower.7ex\hbox{E}\kern-.125emX}}
\usepackage{balance}
\usepackage{epstopdf,amsthm,stfloats,siunitx,amssymb,wasysym,algorithm,algorithmic,array,url, color} 
\begin{document}
\title{Low-Complexity Estimation Algorithm and Decoupling Scheme for FRaC System}
\author{Mengjiang~Sun,~\IEEEmembership{Student~Member,~IEEE,} Peng~Chen,~\IEEEmembership{Senior~Member,~IEEE,} Zhenxin~Cao,~\IEEEmembership{Member,~IEEE,} Fei~Shen,~\IEEEmembership{Member,~IEEE}

 \thanks{This work was supported in part by the Natural Science Foundation for Excellent Young Scholars of Jiangsu Province under Grant BK20220128, the Open Fund of State Key Laboratory of Integrated Chips and Systems under Grant SKLICS-K202305, the Open Fund of National Key Laboratory of Wireless Communications Foundation under Grant IFN20230105, the Open Fund of National Key Laboratory on Electromagnetic Environmental Effects and Electro-optical Engineering under Grant JCKYS2023LD6, the Open Fund of ISN State Key Lab under Grant ISN24-04, and the National Natural Science Foundation of China under Grant 61801112.} 
	\thanks{M. Sun, P. Chen, and Z. Cao are with the State Key Laboratory of Millimeter Waves, Southeast University, Nanjing 210096, China (e-mail: \{mengjiangsun, chenpengseu, caozx\}@seu.edu.cn). Peng~Chen is also with State Key Laboratory of Integrated Chips and Systems, Fudan University, Shanghai 201203, China. F. Shen is with the Key Laboratory of Wireless Sensor Network and Communications, Shanghai Institute of Microsystem and Information Technology, Chinese Academy of Sciences, Shanghai, 200050, China (e-mail: fei.shen@mail.sim.ac.cn)}   \thanks{\textit{(Corresponding author: Peng Chen)}}}

\markboth{IEEE Transactions on Intelligent Vehicles}%
{}

\maketitle

\begin{abstract}
	With the leaping advances in autonomous vehicles and transportation infrastructure, dual function radar-communication (DFRC) systems have become attractive due to the size, cost and resource efficiency. A frequency modulated continuous waveform (FMCW)-based radar-communication system (FRaC) utilizing both sparse multiple-input and multiple-output (MIMO) arrays and index modulation (IM) has been proposed to form a DFRC system specifically designed for vehicular applications. In this paper, the three-dimensional (3D) parameter estimation problem in the FRaC is considered. Since the 3D-parameters including range, direction of arrival (DOA) and velocity are coupled in the estimating matrix of the FRaC system, the existing estimation algorithms cannot estimate the 3D-parameters accurately. Hence, a novel decomposed decoupled atomic norm minimization (DANM) method is proposed by splitting the 3D-parameter estimating matrix into multiple 2D matrices with sparsity constraints. Then, the 3D-parameters are estimated and efficiently and separately with the optimized decoupled estimating matrix. Moreover, the Cram\'{e}r-Rao lower bound (CRLB) of the 3D-parameter estimation are derived, and the computational complexity of the proposed algorithm is analyzed. Simulation results show that the proposed decomposed DANM method exploits the advantage of the virtual aperture in the existence of coupling caused by IM and sparse MIMO array and outperforms the co-estimation algorithm with lower computation complexity.
\end{abstract}

\begin{IEEEkeywords}
Automotive radar, dual function radar communication system, index modulation, atomic norm.
\end{IEEEkeywords}

\section{Introduction}

\IEEEPARstart{R}{ecent} years have witnessed the advances in autonomous vehicles, marked by the incorporation of heightened intelligence ensuring the comfort and safety of advanced driver assistance systems (ADAS), even in intricate and uncontrolled environments~\cite{ADAS_TIV,Automotive_overview,V1,V2}. To fulfill the requirements, vehicles are equipped with a range of sensing technologies including cameras, light detection and ranging (LIDAR) laser-based sensors, global navigation satellite system (GNSS) and automotive radars. Among these sensors, automotive radars are the essential devices due to the capability to detect distant objects in adverse weather conditions and low visibility scenarios~\cite{Automotive}. Beyond environmental sensing, autonomous vehicles are also required to establish potential communication links with neighboring vehicles, passengers, roadside units, and wireless base stations to enable efficient coordination~\cite{C1,C2}. Consequently, autonomous vehicles must be capable of transmitting and processing both radar and communication signals~\cite{dfrc1}.

%% with growing intelligence ensuring the comfort and safety of advanced driver assistance system (ADAS) even in complex uncontrolled environments

Conventional radar and communication systems have historically operated as distinct functionalities, a practice that presents difficulties considering the scarcity of available spectrum resources and the stringent demands placed on system size, cost, and power consumption~\cite{ISAC1}. An alternative strategy is to jointly design both functionalities as a dual function radar-communication (DFRC) system, which has garnered significant attention as a promising technique for autonomous vehicles~\cite{DFRC_TIV,dfrc2,dfrc3,dfrc4,dfrc5,dfrc6}. Existing DFRC schemes for autonomous vehicles can be divided into $4$ main categories as summarized below~\cite{DFRC_vehicle_review}:
\begin{itemize}
	\item \textbf{Coordinated separated waveform schemes}: The probing and communication waveforms are concurrently transmitted with minimal cross-interference in the scheme. This integration can be achieved by leveraging orthogonality in the time or frequency domain, or through the implementation of spatial beamforming techniques~\cite{c1_1,c1_2,c1_3}.
	
	\item \textbf{Communication waveform-based schemes}: The communication waveform serves a dual role as a probing signal in the communication waveform-based DFRC schemes. The most widely adopted choice is orthogonal frequency-division multiplexing (OFDM) waveform due to its spectral efficiency and hardware simplicity. Remarkably, within radar systems, the OFDM waveform circumvents the range-Doppler coupling issue that is typically encountered with frequency-modulated continuous-wave (FMCW) signals~\cite{c2_1,c2_2}. With the above advantages, OFDM waveform becomes a competitive candidate for DFRC schemes. 
	
	%%The advantages in both radar and communication system prove its potential for DFRC systems. However, hardware constraints limits its application. The high peak-to-average power ratio induces distortion in the presence of practical nonlinear amplifiers~\cite{c2_2}.
	\item \textbf{Radar waveform-based schemes}: The communication message is incorporated either within the radar waveform or within the indices of transmission building blocks in the radar waveform-based DFRC schemes. The building blocks encompass spatial allocation, frequency division, and etc~\cite{c2_1,c3_2,c3_3,c3_4}. Embedding information in the indices of transmission building blocks is referred to as index modulation (IM), a technique that has garnered considerable attention within the realm of ongoing DFRC system research~\cite{caesar,majorcom}.
	
	\item \textbf{Joint dual-function waveform schemes}: The joint dual-function waveform is designed 
	to account for the performance of both radar and communications. Through joint optimization, performance trade-off between radar and communications can be achieved at any degree. However, the optimization process can be intricate and multifaceted~\cite{c4_1,c4_2}.
\end{itemize}

%	\begin{table}[h] \caption{DFRC schemes for autonomous vehicles}
	%		\centering
	%		\begin{tabular}{|c|c|} 
		%			\hline
		%			DFRC schemes                        & Explanation                                     \\ \hline
		%			\makecell[c]{Coordinated separated \\ waveform schemes}            & \makecell[c]{ Use different signals for  radar and \\ communication while mitigating the \\ cross interference } \\ \hline
		%			\makecell[c]{Communications \\ waveform schemes}     & \makecell[c]{ Use standard communications signals \\ for probing }            \\ \hline
		%			\makecell[c]{Radar \\ waveform schemes}     & \makecell[c]{ Embed communication message \\ in radar waveforms }            \\ \hline
		%			\makecell[c]{ Joint waveform \\ design schemes}         &  \makecell[c]{ Design the waveform to approach \\ some desired observations at the \\ communication receivers 
			%			and \\ the radar targets }  \\ \hline
		%		\end{tabular}
	%		\label{dfrc}
	%	\end{table}

As mentioned above, the IM technique is popular among recently proposed DFRC systems. The term index can represent the carrier frequency, time slot, antenna allocation or orthogonal waveforms. {Radar performances remain almost unchanged, and the communication rates are contingent on the degree of freedom available}~\cite{frac,c1_3,c3_3,c3_1,ref4,ref5}. 
Multi-carrier agile joint radar communication (MAJoRCom) is proposed to exploit the inherent spatial and spectral randomness of multi-carrier agile phased array radar (CAESAR) to improve the embedded bits~\cite{majorcom,caesar}. The scheme is primarily tailored for detecting remote targets, while in autonomous vehicles, phased array is generally unnecessary due to the proximity of the targets to the radar.  Besides, FMCW waveform better aligns with vehicular system requirements compared to traditional pulse waves, offering established accuracy. In~\cite{frac}, a FMCW-based radar-communications (FRaC) system is proposed as a DFRC system for autonomous vehicles. FRaC extends MAJoRCom to utilize sparse MIMO arrays and FMCW  waveform with modulated phases. In radar performances, the accuracy of detection is improved due to the virtual aperture technique. For communication, FRaC embeds additional phase modulation (PM) symbols in the transmitted waveform since beamforming is unnecessary. Therefore, FRaC becomes an IM-based DFRC system specifically geared towards vehicular applications.

Due to the use of the sparse MIMO array and the IM technique, the target range, DOA and velocity are coupled when processing the signals. The phase differences among signals from adjacent matched filters, neighboring virtual array elements, and successive pulses are influenced by multiple target parameters, which indicates that the parameters cannot be estimated separately with the phase differences. In~\cite{frac}, orthogonal matching pursuit (OMP) algorithm~\cite{OMP} is used for estimating the target parameters in FRaC system. The dictionary is composed of three-dimensional (3D) steering vectors simultaneously concerning the target range, DOA and velocity. To overcome grid-mismatch problem in OMP algorithm, atomic norm minimization (ANM) algorithm can be employed, which optimizes variables in continuous parameter domain with semi-definite programming~\cite{OMP,ANM}. Other algorithms like multiple signal classification (MUSIC)~\cite{MUSIC}, iterative thresholding algorithm (IST)~\cite{IST} and decoupled ANM (DANM)\cite{DANM} are also applicable. All the above algorithms using the 3D steering vectors can be classified as co-estimation algorithms, which estimates the target range, DOA and velocity simultaneously. However, the co-estimation algorithms suffer from high computation burden due to the substantial size of the steering vector. An alternative approach to estimate the coupled parameters is through decoupling algorithms, but the existing decoupling algorithms do not align well with the the estimation requirements in FRaC system. A similar estimation problem is found in FDA-MIMO systems~\cite{FDA_MIMO}, which also leverages spectral freedom in MIMO radar systems but without the use of sparse MIMO arrays and IM techniques. The coupling in FDA-MIMO systems can be eliminated through reshaping the signal matrix, but this approach proves ineffective in the FRaC system~\cite{FDA_MIMO_MUSIC,DANM}. As of now, a decoupling estimation algorithm specifically tailored for the FRaC system has yet to be proposed.
%	
%	while OMP algorithm suffers from high computational complexity and the performance is limited by grid mismatch problems.
%	
%%	Each column in the dictionary matrix is a steering vector concerning the range, DOA and velocity grid point, with the grid spacing equaling the resolution 
%	
%	
%	 and each column vector in the dictionary matrix corresponds to a target range grid, a target DOA grid and a target velocity grid. Difference between adjacent grids equals the resolution. This algorithm cannot exceeds the resolution, and the computational complexity is high. Except for the compressed sensing algorithm, the estimation of target parameters in FRaC system is not further studied. Since the target range, DOA and velocity are coupled when processing the signal, the estimation can only be realized through co-estimation or decoupling the parameters. Computational burden of co-estimation algorithms are relatively high, and existing decoupling algorithms are also not completely suitable for the estimation in FRaC system. The estimation problem in FRaC system is similar to that in FDA-MIMO system, which also utilizes spatial and spectral freedom in MIMO radar system~\cite{FDA_MIMO}. However, FDA-MIMO system does not adopt virtual aperture and index modulation technique as in FRaC system, and thus the coupling in FDA-MIMO system can be eliminated through reshaping the signal matrix, which is not applicable in FRaC system~\cite{FDA_MIMO_MUSIC,DANM}. Therefore, the super-resolution estimation algorithm for FRaC system has not been proposed. 

In this paper, the range-DOA-velocity estimation problem for the FRaC system is considered. We first propose a decoupling scheme, which recovers the decoupled 3D tensor through elements selection and sparsity constraints. To simultaneously limit the sparsity in parameter domain concerning the target range, DOA and velocity, a decomposed DANM (DDANM) algorithm is proposed to form the sparsity constraint and reduce the computational complexity. Through splitting the 3D tensor into multiple 2D matrices and attach sparsity constraint to each of them, DDANM algorithm can efficiently estimate the parameters. Higher-order orthogonal iteration (HOOI) algorithm is then applied on the decoupled 3D tensor to realize Tucker decomposition and estimate the factor matrices. For comparison, we also use the optimized Toeplitz matrix to estimate the target parameters and match the parameters through traversing and projecting. Finally, the 3D-parameters are estimated with the factor matrices using root-MUSIC algorithm. For comparison, we also use the optimized Toeplitz matrix to estimate the target parameters and match the parameters through traversing and projecting. Moreover, the Cram\'{e}r-Rao lower bound (CRLB) of the 3D-parameter estimation are derived, and the computational complexity of the proposed algorithm is analyzed. The main contributions of this paper are summarized below:
\begin{itemize}[itemsep= 0 pt,topsep = -1 pt,partopsep = -5 pt]
	\item \textbf{A decoupling scheme for 3D-parameter estimation}: By introducing a selection matrix, the target parameter coupling is mitigated and the target estimation problem can be transferred into an optimization problem.
	\item \textbf{A decomposed DANM algorithm}: To reduce the computational complexity of 2-fold DANM algorithm, an algorithm based on ANM with decomposed matrices is proposed. The simulation results show that the performance of the proposed algorithm is better than 2-fold DANM in high SNR conditions.
	\item \textbf{The CRLB of the 3D-parameter estimation}: The CRLB for the target parameter estimation in the FRaC system are derived.
	\item \textbf{The application of Tucker decomposition}: HOOI is applied on the recovered 3D tensor to realize Tucker decomposition, which saves the 3D-parameters matching step. 
\end{itemize}

The rest of this paper is organized as follows. In Section~\ref{sec2}, we build the system model for the FRaC system. In Section~\ref{sec3}, the decoupling scheme and the estimation algorithms are presented. The CRLB for the target 3D-parameter estimation is derived in Section~\ref{sec4}. The computational complexity of the algorithms are analyzed in Section~\ref{sec5}. Simulation results and analysis are shown in Section~\ref{sec6}. Finally, Section~\ref{sec7} concludes this paper.

\emph{Notations:} $\boldsymbol{x}^\text{T}$, $\boldsymbol{x}^\text{H}$ and $\boldsymbol{x}^\text{*}$ denote the transpose, the Hermitian transpose and the conjugation of $\boldsymbol{x}$, respectively. $\| \boldsymbol{x} \|_1$ and $\| \boldsymbol{x} \|_2$ denote the $\ell_1$ and $\ell_2$ norm of $\boldsymbol{x}$, respectively. $\mathrm{Tr} \{ \boldsymbol{X} \}$ denotes the trace of $\boldsymbol{x}$. $\mathbb{E} \{ \boldsymbol{x} \}$ denotes the expectation of $\boldsymbol{x}$. $\mathcal{R} \{a\}$ denotes the real part of complex value $a$. $\otimes$ denotes the Kronecker product. $\odot$ denotes the Khatri-Rao product. The boldface capital letters denote
the matrix, such as $\boldsymbol{X}$, and the lower-case letters denote the vector, such as $\boldsymbol{x}$. The n-mode product is written as $\times_n$. The outer product is written as $\circ$.

\section{The FRaC System Model} \label{sec2}

\begin{figure}[h] 
	\centering
	\includegraphics[width=0.475\textwidth]{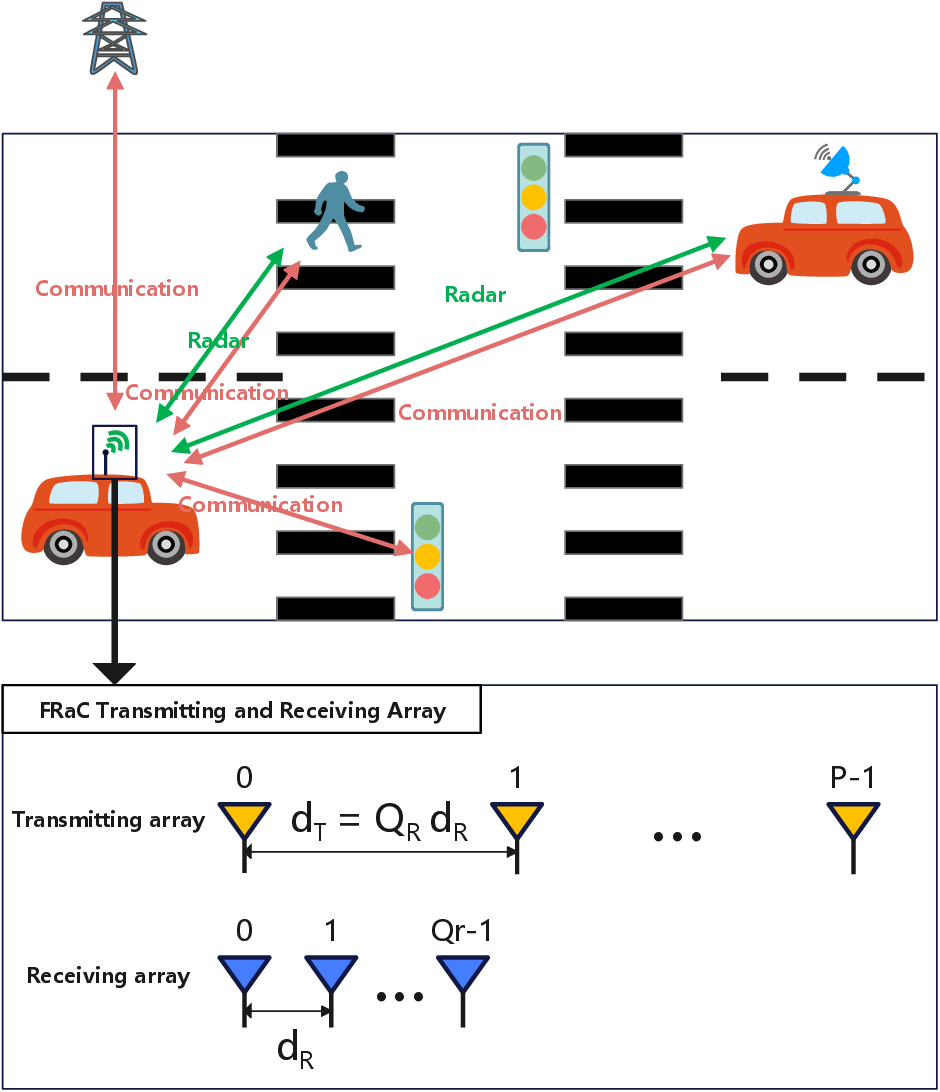}
	\caption{A diagram of the DFRC system for vehicular applications.}
	\label{sketch}
\end{figure}

As shown in Fig.~\ref{sketch}, the DFRC system for vehicular applications includes various types of communication and radar detection, including communication with passengers, wireless base stations, road-side units and radar detection targeting passengers, other vehicles and etc. The communication link is represented with red lines, and the transmitting and reflecting of the probing waves are shown with greeen lines. We then show the system model of the FRaC system, together with the preliminary radar signal processing.

\subsection{Transmission Model of FRaC}

The FRaC system utilizes a ULA with $P$ elements to emit FMCW signal at the transmitter. During each transmission, $K$ FMCW waveforms with different carrier frequencies are multiplied with $K$ phase modulation (PM) symbols, respectively~\cite{frac}. In the $n$-th pulse, $n \in [0,1,\dots,N-1]$, the transmitted waveform assigned to the $k$-th active element is
\begin{equation}
	x_{n,k} (t) = s(t-nT_0) e^{j 2 \pi f_{n,k}t} e^{j \phi_{n,k}},
\end{equation}
where $f_{n,k} = f_c + m_{n,k} \Delta f$. $f_c$ is the start of carrier frequency. $m_{n,k} \in [0,1,\cdots,M-1]$ and $M$ denotes the number of optional frequency points. $\phi_{n,k}$ is the PM symbol modulated on the waveform. $s(t)$ is the baseband FMCW wave, which is expressed as
\begin{equation}
	s(t) = \mathrm{rect} \left( \frac{t}{T_0} \right) e^{j k_f \pi t^2}, 0 \leq t \leq T_0,
\end{equation}
where $\mathrm{rect} (t) = 1$ for $0 \leq t \leq 1$ and $T_0$ denotes the duration of one FMCW pulse.

Assume that each PM symbol takes $J$ possible values, the number of bits conveyed by PM during each transmission is $N_{\text{PM}} = K \mathrm{log_2} J$, with $\mathrm{log_2}$ denoted as the logarithm to the base 2. The frequencies of the $K$ transmitted waveform are selected from $M$ possible frequencies, which embeds $\mathrm{log_2} {M \choose K} = \mathrm{log_2} \frac{M!}{(M-K)!K!}$ bits with the selection. Additionally, selecting $K$ out of $P$ antenna elements for transmission enables the embedding of $\mathrm{log_2} {M \choose K}$ bits. Finally, assigning the selected $K$ waveform to the $K$ selected elements embeds $\mathrm{log_2} K!$ bits. Hence, the total number of bits conveyed in each pulse is $K \mathrm{log_2} J+\mathrm{log_2} {M \choose K}+\mathrm{log_2} {M \choose K}+\mathrm{log_2} K!$.

\subsection{Radar Signal Processing}

Consider a scenario where there are $L$ targets situated in the far field, each characterized by parameters including range ${ r_l }$, velocity ${v_l }$, and DOA ${\theta_l }$. $p_{n,k} \in [0,1,\cdots,P-1]$ denotes the index of the $k$-th active antenna in the $n$-th pulse. The round-trip delay between the $p_{n,k}$-th transmit element and the $q_r$-th receive element in the $n$-th pulse is 
\begin{equation}
	\tau_{n,k,q_r}^{l} (t) = \frac{2 ( r_l + v_l t)}{c} - \frac{(p_{n,k} d_T+q_r d_R) \sin{\theta_l}}{c},
\end{equation}
where $q_r \in [0,1,\cdots,Q_r - 1]$ and $Q_R$ denotes the number of receiving antennas. $t \in \left[ (n-1)T_0, n T_0 \right]$. $d_T$ and $d_R$ denote distance between adjacent elements in transmitting and receiving array, respectively. Since virtual aperture technique is used in the scheme, we have $d_T = Q_r d_R$. $c$ is the speed of light. The received signal at the $q_r$-th element is 
\begin{equation}
	y_{n,q_r} (t) = \sum_{l=0}^{L-1} \alpha_l \sum_{k=0}^{K-1} x_{n,k}(t-\tau_{n,k,q_r}^{l} (t)) + w_{n,q_r}.
\end{equation}
$\alpha_l$ is the reflective factor of the $l$-th target. The received signal is then mixed with the conjugate of the transmitting signal $x_{n,k}(t)$ and then fed into a low pass filter to extract the desired signals, which is expressed as 
\begin{equation}
	\begin{split}
		y_{n,k,q_r}^{(r)} (t) &  = \mathrm{LPF} ( y_{n,q_r} (t) x_{n,k}^*(t))
		\\ & =\sum_{l = 0} ^ {L-1} \tilde{\alpha}_l  e^{-j 2 \pi k_f \tau_{n,k,q_r}^{l}(t) (t-nT_0)}  e^{-j 2 \pi m_{n,k} \Delta f \frac{2 r_l}{c}} 
		\\ & \quad  e^{-j 2 \pi f_c \frac{2 \pi v_l T_0 - ( p_{n,k} d_T + q_r d_R) \sin ( \theta_l )}{c} } + \tilde{w}_{n,k,q_r}(t),
	\end{split}
\end{equation}
where $\tilde{\alpha}_l \triangleq e^{-j 2 \pi f_c \frac{2 r_l}{c} }$. $\mathrm{LPF}(\boldsymbol{x})$ denotes the output of the low pass filter with the input $\boldsymbol{x}$. The equivalent band-limited Gaussian noise is denoted as $\tilde{w}_{n,k,q_r}(t) \triangleq \mathrm{LPF} ( w_{n,q_r} (t) x_{n,k}^*(t))$. Assume that the sample time instances are $t = nT0 + \tilde{g} T_s, \tilde{g} = 0,1,\cdots,G-1$, where $T_s$ is the sampling interval and $G$ is the number of sample points in each pulse repetition interval (PRI). The sampled signal is 
\begin{equation}
	\begin{split}
		& y_{n,k,q_r}^{(r)} [g] =\sum_{l = 0} ^ {L-1} \tilde{\alpha}_l  e^{-j 2 \pi k_f \left( \frac{2 ( r_l + v_l n T_0)}{c} - \frac{(p_{n,k} d_T+q_r d_R) \sin{\theta_l}}{c} \right) g T_s}   
		\\ & e^{-j 2 \pi \left( m_{n,k} \Delta f \frac{2 r_l}{c} + f_c \frac{2 \pi v_l T_0 - ( p_{n,k} d_T + q_r d_R) \sin ( \theta_l )}{c} \right) } + \tilde{w}_{n,k,q_r}(t),
	\end{split}
\end{equation}

We then generate the coarse range profile $\hat{y}_{n,k,q_r}^{(r)} [g]$ with discrete Fourier transform (IDFT),

\begin{equation}
	\begin{split}
		& \hat{y}_{n,k,q_r}^{(r)} [g]  = \mathrm{IDFT} (y_{n,k,q_r}^{(r)} [g]) = \sum_{\tilde{g}=0}^{G-1} y_{n,k,q_r}^{(r)} [\tilde{g}] e^{\frac{j 2 \pi \tilde{g} g}{G}}
		\\ & = \sum_{l = 0} ^ {L-1} \beta_l [g]  e^{-j 2 \pi f_c \frac{2 n v_l T_0 - ( p_{n,k} d_T + q_r d_R) \sin ( \theta_l ) + 2 m_{n,k} \Delta f  r_l}{c} } 
		\\ & \quad + \hat{w}_{n,k,q_r} [g],
	\end{split}
\end{equation}
where $\mathrm{IDFT}(\cdot)$ denotes the IDFT operation over $G$ samples.
$\beta_l [g] \triangleq \tilde{\alpha}_l \sum_{\tilde{g}=0}^{G-1} e^{j 2 \pi \tilde{g} (\frac{g}{G} - \frac{2 k_f r_l T_s}{c})}$, and  $\hat{w}_{n,k,q_r} [g]$ is the additional Gaussian noise.

$\hat{y}_{n,k,q_r}^{(r)} [g]$ refers to the $g$-th coarse resolution range profile with range resolution $\frac{c}{2 k_f T_p}$. The 3D-parameters of the targets within the $g$-th resolution range are estimated with $\hat{y}_{n,k,q_r}^{(r)} [g]$, and samples from other coarse range cells are processed identically and separately.

%% Since $nT_0 v_l$, $d_T$ and $d_R$ are far less than $r_l$, we have ${\tau_{n,k,q_r}^{l}}^2 \approx \frac{4 r_l^2}{c^2}$. $\beta_l$ is then simplified as $\tilde{\alpha}_l e^{j \pi k_f \frac{4 r_l^2}{c^2}}$.

We then reshape the $3$-order tensor $ \hat{\boldsymbol{y}}^{(r)} [g] \in \mathbb{C}^{N \times K \times Q_r}$ into a matrix $\boldsymbol{Y} \in \mathbb{C}^{NK \times Q_r}$. $\boldsymbol{Y}$ can be represented as
\begin{equation} \label{model}
	\boldsymbol{Y} =\sum_{l=1}^{L} \beta_l \left [\boldsymbol{a}_r (r_l) \odot \boldsymbol{a}_{v} (v_l) \odot \boldsymbol{a}_{t} (\theta_l) \right ] \boldsymbol{b}_{r} (\theta_l)^T + \boldsymbol{\hat{w}},
\end{equation}
where $\boldsymbol{\hat{w}}$ denotes the additional Gaussian white noise. $\boldsymbol{a}_r(r) \in \mathbb{C}^{NK \times 1}$ denotes the steering vector concerning the range of the target. $\boldsymbol{a}_v(v) \in \mathbb{C}^{NK \times 1}$ denotes the steering vector concerning the velocity of target. $\boldsymbol{a}_t(\theta) \in \mathbb{C}^{NK \times 1}$ denotes the transmitting steering vector concerning the DOA of target. $\boldsymbol{b}_r(\theta) \in \mathbb{C}^{NK \times 1}$ denotes the receiving steering vector concerning the DOA of target. The $\left [(n-1)K+k \right ]$-th entry of $\boldsymbol{a}_r(r)$, $\boldsymbol{a}_v(v)$, $\boldsymbol{a}_t(\theta)$ and the $q_r$-th entry of $\boldsymbol{b}_r(\theta)$ are

\begin{equation}
\begin{aligned} 
	{a}_{r,(n-1)K+k}(r)  & = e^{-j 2 \pi m_{n,k} \Delta f \frac{2 r}{c}},
        \\{a}_{v,(n-1)K+k}(v) &= e^{-j 2 \pi f_c \frac{2 n v T_0}{c}},
	\\{a}_{t,(n-1)K+k}(\theta) &= e^{j 2 \pi p_{n,k} f_c \frac{d_T \sin(\theta)}{c}},
	\\{b}_{r,q_r}(\theta) &= e^{j 2 \pi q_r f_c \frac{d_R \sin(\theta)}{c}}.
\end{aligned}
\end{equation}

\section{Estimation of 3D-Parameter} \label{sec3}                

\subsection{ Decoupling of Target Parameters}

According to the structure of $\boldsymbol{Y}$ in (\ref{model}), the target range $r_l$, the target velocity $v_l$ and the target DOA $\theta_l$ are coupled in the columns of $\boldsymbol{Y}$. In each row of $\boldsymbol{Y}$, phase differences exclusively concern with $\theta_l$. Nevertheless, if $\theta_l$ is directly estimated with the rows of $\boldsymbol{Y}$, the potential advantage of virtual aperture technology remains unexploited, resulting in a limitation on the number of distinguishable targets and a degraded precision of the estimated DOA. To overcome the problems, we propose a method to fully utilize the virtual aperture and estimate the parameters without loss of accuracy. 

\begin{figure}[h] 
	\centering
	\includegraphics[width=0.475\textwidth]{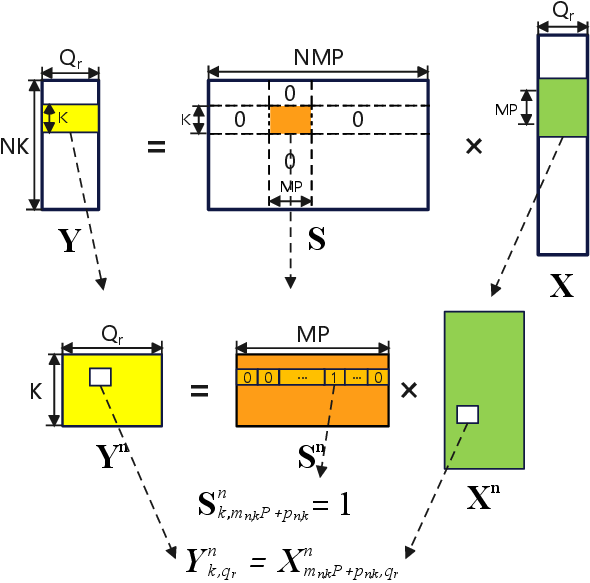}
	\caption{A diagram illustrating the relation among the matrix $\boldsymbol{Y}$, the selection matrix $\boldsymbol{S}$ and the reference matrix $\boldsymbol{X}$.}
	\label{selection}
\end{figure}

We first introduce a selection matrix $\boldsymbol{S} \in \mathbb{R} ^ {NK \times NMP}$, with its $(nK+k)$-th row, the $(nMP + mP + p)$-th column denoted as
\begin{equation}
	\boldsymbol{S}_{nK+k,nMP + mP + p} = \left\{\begin{matrix} 1, \; p_{n,k} = p, m_{n,k} = m,
		\\0, \; \mathrm{else}.\qquad \qquad \qquad \: \: \;
	\end{matrix}\right.
\end{equation}
The matrix $\boldsymbol{Y}$ can be then represented with $\boldsymbol{S}$ as 

\begin{equation}
	\boldsymbol{Y} = \boldsymbol{S} \boldsymbol{X}.
\end{equation}
$\boldsymbol{X} \in \mathbb{C}^{MNP \times Q_r}$ represents a reference matrix, which includes all possible combinations of active transmitting antennas and active frequencies. For example, the first row of $\boldsymbol{X}$ denotes the expected signal when only the first antenna element is activated, which transmits signal with a frequency of $f_c$. {The relation among the matrix $\boldsymbol{Y}$, the selection matrix $\boldsymbol{S}$ and the reference matrix $\boldsymbol{X}$ is shown in Fig.}~\ref{selection}. {$\boldsymbol{Y}^n$, $\boldsymbol{S}^n$, $\boldsymbol{X}^n$ denote the submatrices concerning the $n$-th pulse in $\boldsymbol{Y}$, $\boldsymbol{S}$, $\boldsymbol{X}$, respectively. In the $k$-th row of the sub-selection matrix, the $(m_{n,k}P+p_{n,k})$-th element equals $1$ and the other elements equal $0$. Such design of the selection matrix ensures that all elements in matrix $\boldsymbol{Y}$ are correctly projected to the corresponding elements in the reference matrix $\boldsymbol{X}$.}

{The reference matrix $\boldsymbol{X}$ can be reshaped into a $3$-order tensor with the elements along each dimension only varying with one parameter, suggesting a decoupled structure.} $\boldsymbol{X}$ is represented as
\begin{equation}
	\boldsymbol{X} = \sum_{l=1}^{L}  \beta_l \left [\boldsymbol{\tilde{a}}_r (r_l) \otimes \boldsymbol{\tilde{a}}_{v} (v_l) \otimes \boldsymbol{\tilde{a}}_{t} (\theta_l) \right ] \boldsymbol{\tilde{b}}_{r} (\theta_l)^T,
\end{equation}
where $\boldsymbol{\tilde{a}}_r (r)$, $\boldsymbol{\tilde{a}}_v (v)$, $\boldsymbol{\tilde{a}}_t (\theta)$ and $\boldsymbol{\tilde{b}}_r (\theta)$ are denoted as 
\begin{equation}
\begin{aligned}
	\boldsymbol{\tilde{a}}_r (r) &= e^{-j 2 \pi [0,1, \cdots, M-1]^{T} \Delta f \frac{2 r}{c}},
	\\ \boldsymbol{\tilde{a}}_v (v) &= e^{-j 2 \pi [0,1, \cdots, N-1]^{T} f_c \frac{2 nvT_0}{c}},
	\\ \boldsymbol{\tilde{a}}_t (\theta) &= e^{j 2 \pi [0,1, \cdots, P-1]^{T} f_c \frac{d_T \sin (\theta)}{c}},
	\\ \boldsymbol{\tilde{b}}_r (\theta) &= e^{j 2 \pi [0,1, \cdots, Q_r-1]^{T} f_c \frac{d_R \sin (\theta)}{c}}.
\end{aligned}
\end{equation}
Therefore, the decoupling of the target range, the target velocity and the target DOA can be achieved through reshaping $\boldsymbol{X}$. The subsequent objectives include recovering $\boldsymbol{X}$ with $\boldsymbol{Y}$ and estimating the target parameters with the recovered $\boldsymbol{X}$.

\subsection{ Estimation Based on 2-fold DANM Algorithm }

\begin{figure}[h] 
	\centering
	\includegraphics[width=0.475\textwidth]{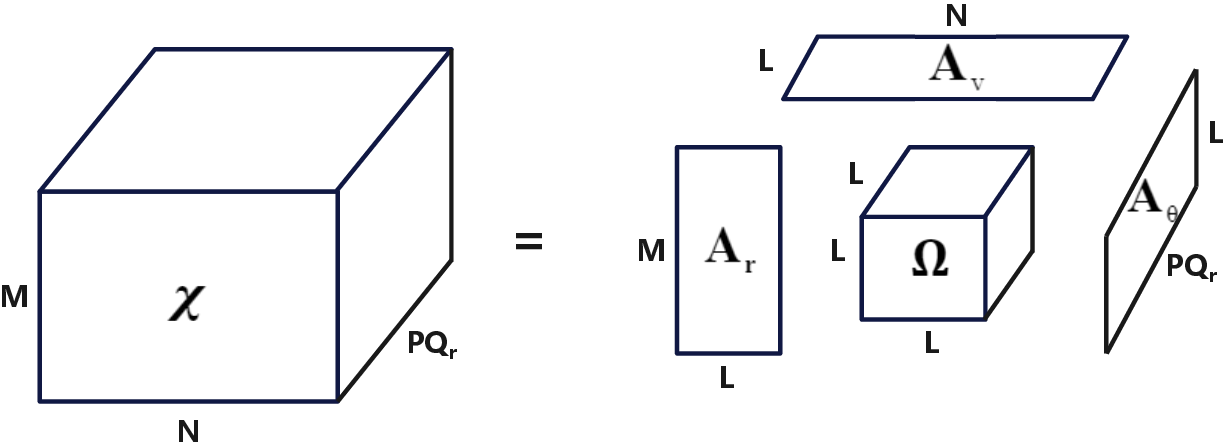}
	\caption{Diagram of Tucker decomposition.}
	\label{Tucker}
\end{figure}

We first reshape $\boldsymbol{X}$ into $\boldsymbol{X'} \in \mathbb{C}^{MN \times PQ_r}$ ,which satisfies 
\begin{equation}
	\boldsymbol{X'}_{(m-1)N+n,(p-1)Q_r+q} = \boldsymbol{X}_{(m-1)NP + (n-1)P +p,q}.
\end{equation}
Virtual aperture is fully exploited in each row of $\boldsymbol{X'}$, which only concerns with the target DOA. Since the column vectors in $\boldsymbol{X'}$ consist of steering vectors concerning both the target range and the target velocity, we use 2-fold DANM to obtain subspace concerning the target DOA and the target velocity~\cite{ANM,ANM_2d}. The problem can be written as
\begin{equation} \label{2-fold}
	\begin{aligned}\min_{\boldsymbol{X},\boldsymbol{T},\boldsymbol{u}} &  \left \| \boldsymbol{Y} - \boldsymbol{S} \boldsymbol{X} \right \|_F^2 + \tau ( \mathrm{Tr} (\mathrm{S}(\boldsymbol{T})) + \mathrm{Tr} ( \mathrm{T} (\boldsymbol{u}) ) ) \\
		\text { s.t. } &\left[\begin{array}{cc}
			\mathcal{S}(\boldsymbol{T}) & \boldsymbol{X'} \\
			\boldsymbol{X'}^{\text{H}} & \mathrm{T} (\boldsymbol{u})
		\end{array}\right] \succeq 0
	\end{aligned}
\end{equation}
$\tau$ is the regularization parameter, which determines the strength of the sparsity constraint. $\mathrm{S} ( \boldsymbol{T} ) \in \mathbb{C}^{NM \times NM}$ is a two-fold Toeplitz matrix consisting of $N \times N$ block Toeplitz matrices, i.e.,
\begin{equation}
	\mathcal{S}(\boldsymbol{T})=\left[\begin{array}{cccc}
		\boldsymbol{T}_0 & \boldsymbol{T}_{-1} & \cdots & \boldsymbol{T}_{-\left(M-1\right)} \\
		\boldsymbol{T}_1 & \boldsymbol{T}_0 & \cdots & \boldsymbol{T}_{-\left(M-2\right)} \\
		\vdots & \vdots & \vdots & \vdots \\
		\boldsymbol{T}_{M-1} & \boldsymbol{T}_{M-2} & \cdots & \boldsymbol{T}_0
	\end{array}\right].
\end{equation}
Block $\boldsymbol{T}_l \in \mathbb{C}^{N \times N}$ is defined as 
\begin{equation}
	\boldsymbol{T}_l=\left[\begin{array}{cccc}
		x_{l, 0} & x_{l,-1} & \cdots & x_{l,-\left(N-1\right)} \\
		x_{l, 1} & x_{l, 0} & \cdots & x_{l,-\left(N-2\right)} \\
		\vdots & \vdots & \vdots & \vdots \\
		x_{l, N-1} & x_{l, N-2} & \cdots & x_{l, 0}
	\end{array}\right],
\end{equation}
$\mathrm{T} (\boldsymbol{u})$ represents the Toeplitz matrix constructed by $\boldsymbol{u}$. After solving the convex optimization problem, we reshape the recovered $\boldsymbol{X}$ into a 3-order tensor $\boldsymbol{\chi} \in \mathbb{C}^{M \times N \times PQ_r}$, which satisfies  
\begin{equation}
	\begin{split}
		\boldsymbol{\chi} & = \sum_{l=1}^{L}  \beta_l \circ \boldsymbol{\tilde{a}}_{v} (v_l) \circ \boldsymbol{\tilde{a}}_r (r_l) \circ \left( \boldsymbol{\tilde{a}}_{t} (\theta_l) \otimes \boldsymbol{\tilde{b}}_{r} (\theta_l) \right)
		\\ & = \boldsymbol{\Omega} \times_1 \boldsymbol{{A}}_{v} \times_2 \boldsymbol{{A}}_{r} \times_3 \boldsymbol{{A}}_{\theta}.
	\end{split}
\end{equation}
$\boldsymbol{\Omega} \in \mathbb{C}^{L \times L \times L}$ denotes the core tensor. $\boldsymbol{A}_v \in \mathbb{C}^{N \times L}$ denotes the factor matrix concerning the target velocity, which is composed of the steering vectors $\boldsymbol{\tilde{a}}_{v} (v_l),l=1,2,\cdots,L$. Similarly, $\boldsymbol{A}_r$ and $\boldsymbol{A}_{\theta}$ denote the factor matrices concerning the target range and the target DOA, respectively. We then adopt Tucker decomposition to decompose $\boldsymbol{\chi}$ and estimate the factor matrices. An illustration of Tucker decomposition is shown in \cref{Tucker}.

An advantage of utilizing Tucker decomposition is saving the parameters matching step. When estimated separately, the target ranges, DOAs and velocities require matching, and the computational burden is high when the number of targets is large. With Tucker decomposition, the target parameters matching is completed through obtaining the index of non-zero elements in $\boldsymbol{\Omega}$. For example, the non-zero element $\Omega_{i,j,k}$ indicates that the $i$-th column vector in $\boldsymbol{A}_v$, the $j$-th column vector in $\boldsymbol{A}_r$ and the $k$-th column vector in $\boldsymbol{A}_{\theta}$ correspond to the same target.

HOOI algorithm and higher-order singular value decomposition (HOSVD) algorithm are known as the gold standard for computing the factor matrices of a Tucker Decomposition \cite{HOOI,rankadaptive,HOSVD}. Since HOOI algorithm usually gives better results of the best rank approximation compared to HOSVD algorithm \cite{Tucker}, we use the result of HOSVD as an initialization for HOOI algorithm. After obtaining the factor matrices, we use root-MUSIC algorithm to estimate the 3D target parameters. The details of the proposed algorithm is given in \cref{algorithm1}.

\begin{algorithm}
	\renewcommand{\algorithmicrequire}{\textbf{Input:}}
	\renewcommand{\algorithmicensure}{\textbf{Output:}}
	\caption{ Estimation based on 2-fold DANM algorithm and HOOI algorithm }
	{\label{algorithm1}}
	\begin{algorithmic}[1]
		\REQUIRE The signal matrix $\boldsymbol{{Y}}$, the selection matrix $\boldsymbol{S}$, the regularization parameter $\tau$.
		\ENSURE The target range $\boldsymbol{r}$, the target velocity $\boldsymbol{v}$, the target DOA $\boldsymbol{\theta}$.
		\STATE Solve the problem (\ref{2-fold}) via convex optimization.
		\STATE Use HOOI algorithm to realize Tucker decomposition for the optimized decoupled tensor $\tilde{\boldsymbol{X}}$.
		\STATE The target range $\boldsymbol{r}$, velocity $\boldsymbol{v}$ and DOA $\boldsymbol{\theta}$ are estimated through applying root-MUSIC algorithm on the decomposed factor matrices.
		\STATE \textbf{return} $\boldsymbol{r}$, $\boldsymbol{v}$, $\boldsymbol{\theta}$.
	\end{algorithmic}  
\end{algorithm} 	

	An alternative method to estimate the 3D target parameters after optimization (\ref{2-fold}) is decomposing the optimized Toeplitz matrices $\boldsymbol{T}_l$ and $\mathrm{T} (\boldsymbol{u})$. The covariance matrix concerning the target velocity and DOA are represented as $\boldsymbol{R}_v$ and $\boldsymbol{R}_u$, respectively, i.e.
	\begin{equation} \label{R1}
			\begin{aligned}
					& \boldsymbol{R}_v = \sum_{m = 0}^{M-1} \mathrm{T}_m \mathrm{T}_m^H,
					\\ & \boldsymbol{R}_u = \mathrm{T} (\boldsymbol{u}) \mathrm{T} (\boldsymbol{u})^H.
				\end{aligned}
		\end{equation}
	The steering vector concerning the target DOA can be obtained from the decomposition of $\boldsymbol{R}_u$, and the steering vector concerning the target velocity can be obtained from decomposition of $\boldsymbol{R}_v$. Thereafter, we use MUSIC algorithm to estimate the target DOA and the target velocity.

	Finally, we match the estimated target DOAs and velocities and then estimate the target ranges. Assume that the estimated target DOA set is $ \hat {\boldsymbol{\theta}}$, and the estimated target velocity set is $ \hat {\boldsymbol{v}}$. The matching problem and the estimation of the target ranges can be formulated as
	\begin{equation} \label{prj1}
			\min_{\hat{r}_l, \hat{\theta}_l \in \hat {\boldsymbol{\theta}}, \atop \hat{v}_l \in \hat {\boldsymbol{v}}, \beta_l} \bigg\| \boldsymbol{Y} - \sum_{l=1}^{L} \beta_l \left [\boldsymbol{a}_{r} (\hat{r}_l) \odot \boldsymbol{a}_{v} (\hat{v}_l) \odot \boldsymbol{a}_{t} (\hat{\theta}_l) \right ] \boldsymbol{b}_{r} (\hat{\theta}_l)^T  \bigg\|_F^2.
		\end{equation}
	We solve the problem through traversing all possible combinations of the estimated target velocities within the set $\boldsymbol{\hat{v}}$ and the estimated target DOAs within the set $\boldsymbol{\hat{\theta}}$. The received signal is then projected onto the space constructed by each combination. With the projected signal, we can estimate the target range for each combination of the estimated target velocity and the estimated target DOA. The residual is then obtained for each combination. We finally determine the parameters of estimated targets with the combination corresponding to the minimum residual. The algorithm based on 2-fold DANM algorithm and matching is given in \cref{algorithm2}.
	
	\begin{algorithm}
		\renewcommand{\algorithmicrequire}{\textbf{Input:}}
		\renewcommand{\algorithmicensure}{\textbf{Output:}}
		\caption{ Estimation based on 2-fold DANM algorithm and parameters matching }
		{\label{algorithm2}}
		\begin{algorithmic}[1]
			\REQUIRE The signal matrix $\boldsymbol{{Y}}$, the regularization parameter $\tau$.
			\ENSURE The target range $\boldsymbol{r}$, the target velocity $\boldsymbol{v}$, the target DOA $\boldsymbol{\theta}$.
			\STATE Solve the problem (\ref{2-fold}) via convex optimization.
			\STATE Construct $\boldsymbol{R}_u$ and $\boldsymbol{R}_v$ via equation (\ref{R1}) and apply MUSIC algorithm on $\boldsymbol{R}_u$ and $\boldsymbol{R}_v$ to obtain the estimated target DOA set $\boldsymbol{\hat{\theta}}$ and the estimated target velocity set $\boldsymbol{\hat{v}}$.
			\STATE Solve problem (\ref{prj1}) via projecting $\boldsymbol{{Y}}$ on the signal space constructed by each combination of $ \theta \in \boldsymbol{\hat{\theta}}$ and $ v \in \boldsymbol{\hat{v}}$.
			\STATE The estimated target range $\boldsymbol{r}$, velocity $\boldsymbol{v}$ and DOA $\boldsymbol{\theta}$ are set as the combination corresponding to the minimum residual.
			\STATE \textbf{return} $\boldsymbol{r}$, $\boldsymbol{v}$, $\boldsymbol{\theta}$.
		\end{algorithmic}  
	\end{algorithm}

\subsection{ Estimation Based on Decomposed DANM Algorithm }

When solving the convex optimization problem with interior point method, it is necessary to take the second derivatives of all variables to derive the Hessian matrix, where the semi-definite constraint is transferred into a determinant constraint. Taking derivatives will be computationally complex when the rank of the determinant is high. Therefore, inspired by decoupled DANM~\cite{DANM}, we propose decomposed DANM (DDANM), which splits the 2-fold Toeplitz matrix into several 1-fold Toeplitz matrix. With DDANM, the problem is reformulated as
\begin{equation} \label{opt}
	\begin{aligned}
		& \min_{\boldsymbol{X},\boldsymbol{u}_m,\boldsymbol{v}_m} \left \| \boldsymbol{Y} - \boldsymbol{S} \boldsymbol{X} \right \|_F^2 + \tau \sum_{m = 1}^{M} \mathrm{Tr} (\mathrm{T}(\boldsymbol{u}_m)+\mathrm{T}(\boldsymbol{v}_m)) 
		\\ & \qquad  s.t. \begin{bmatrix} \mathrm{T} (\boldsymbol{v}_m)   & \boldsymbol{X}_m\\  \boldsymbol{X}_m^H 
			& \mathrm{T} (\boldsymbol{u}_m)\end{bmatrix} \geq 0, m = 1,2, \cdots, M.
	\end{aligned}
\end{equation}
$\mathrm{T} (\boldsymbol{u})$ denotes a Toeplitz matrix constructed by $\boldsymbol{u}$. $\boldsymbol{X}_m \in \mathbb{C}^{N \times PQ_r}$ denotes the matrix concerning $(m-1) \Delta f$ in $\boldsymbol{X}$, i.e., the elements from the $(m-1)N+1$-th row to the $mN$-th row in $\boldsymbol{X}$. We use an extracting matrix $\boldsymbol{H}_m \in \mathbb{R}^{N \times MN}$ to extract the rows concerning $(m-1) \Delta f$ in $\boldsymbol{X'}$. $\boldsymbol{H}_m$ satisfies $\boldsymbol{H}_m \triangleq \bigg [ \boldsymbol{H}_{m,1},\boldsymbol{H}_{m,2},\cdots,\boldsymbol{H}_{m,M} \bigg ]$, with $\boldsymbol{H}_{m,i} \in \mathbb{R}^{N \times N} , i \in 1,2,\cdots,M$. $\boldsymbol{H}_{m,m}$ is an identity matrix and $\boldsymbol{H}_{m,i}, i \neq m$ is a zero matrix. 
With the above definition, $\boldsymbol{X}_m$ can be written as
\begin{equation}
	\boldsymbol{X}_m = \boldsymbol{H}_m \boldsymbol{X'}.
\end{equation}
After solving the problem (\ref{opt}) with semi-definite programming, we can then reshape the optimized $\boldsymbol{X}$ into a 3-order tensor $\tilde{\boldsymbol{X}} \in \mathbb{C}^{M \times N \times PQ_r}$. Tucker decomposition based on HOOI algorithm is then utilized to compute the factor matrices and root-MUSIC algorithm is used to estimate the 3D target parameters. 

	Similar to algorithm~\ref{algorithm2}, we can also estimate and match the 3D-parameters separately. We obtain $\boldsymbol{R}_u$ and $\boldsymbol{R}_v$ with the optimized $\boldsymbol{u}_m$ and $\boldsymbol{v}_m$ as
	\begin{equation} \label{RR}
			\begin{aligned}
					& \boldsymbol{R}_u = \sum_{m = 1}^{M} \mathrm{T} (\boldsymbol{u}_m) \mathrm{T} (\boldsymbol{u}_m)^H,
					\\ & \boldsymbol{R}_v = \sum_{m = 1}^{M} \mathrm{T} (\boldsymbol{v}_m) \mathrm{T} (\boldsymbol{v}_m)^H.
				\end{aligned}
		\end{equation}
	We then estimate the target DOA and the target velocity with $\boldsymbol{R}_u$ and $\boldsymbol{R}_v$.The matching of the target parameters and the estimation of the target range are also the same as in algorithm~\ref{algorithm2}.

\subsection{ Estimation Based on $\ell_1$ Norm Minimization Algorithm }
We also propose an iterative optimization algorithm based on $\ell_1$ norm minimization to further reduce computational complexity. The sparsity in all dimensions including the target range, the target velocity and the target DOA can be represented as the sparsity exhibited in the projection vectors when projecting onto each dictionary matrix. The problem can be written as
\begin{equation} \label{l1}
	\begin{aligned}\min_{\boldsymbol{X},\boldsymbol{T},\boldsymbol{P}} &  \left \| \boldsymbol{Y} - \boldsymbol{S} \boldsymbol{X} \right \|_F^2 + \tau_1 \left \| \boldsymbol{X}_1 - \boldsymbol{A}_1 \boldsymbol{x}_1 \right \|_F^2 + 
		\\ &  \tau_2 \left \| \boldsymbol{X}_2 - \boldsymbol{A}_2 \boldsymbol{x}_2 \right \|_F^2 + \tau_3 \left \| \boldsymbol{X}_3 - \boldsymbol{A}_3 \boldsymbol{x}_3 \right \|_F^2 + \\
		& \mu_1 \left \| \boldsymbol{x}_1 \right \|_1 + \mu_2 \left \| \boldsymbol{x}_2 \right \|_1 + \mu_3 \left \| \boldsymbol{x}_3 \right \|_1
	\end{aligned}
\end{equation}
$\boldsymbol{X}_1 \in \mathbb{C}^{PQ_r \times MN}$ denotes the reshaped $\boldsymbol{X}$ with each column of $\boldsymbol{X}_1$ only concerns with the target DOA. Similarly, we obtain $\boldsymbol{X}_2 \in \mathbb{C}^{M \times NPQ_r}$ and $\boldsymbol{X}_3 \in \mathbb{C}^{N \times MPQ_r}$ with each column vector only concerning with the target range and the target velocity, respectively. $\boldsymbol{A}_1$, $\boldsymbol{A}_2$ and $\boldsymbol{A}_3$ denote the dictionary matrix concerning the target DOA, the target range and the target velocity, respectively, with $\boldsymbol{x}_1$, $\boldsymbol{x}_2$ and $\boldsymbol{x}_3$ denoted as the corresponding projection vector. $\tau_i$ and $\mu_i, i=1,2,3$, represent the regularization parameters. 

Although the optimization problem (\ref{l1}) is also convex, it is hard to directly obtain the optimal solution due to the high dimension of the variables. Therefore, we adopt an iterative approach for the problem. We first project the matrix onto each dictionary to obtain the sparse vector $\boldsymbol{x}_1$, $\boldsymbol{x}_2$, $\boldsymbol{x}_3$. The update for matrix $\boldsymbol{X}$ can be then resolved through quadratic programming. After the estimation, we subsequently match the 3D target parameters. Assume that the estimated target DOA set is $ \hat {\boldsymbol{\theta}}$, and the estimated target velocity set is $ \hat {\boldsymbol{v}}$. The matching problem and the estimation of the target ranges can be formulated as
\begin{equation} \label{prj2}
	\min_{\hat{r}_l  \in \hat {\boldsymbol{r}}, \hat{\theta}_l \in \hat {\boldsymbol{\theta}}, \atop \hat{v}_l \in \hat {\boldsymbol{v}}, \beta_l} \bigg\| \boldsymbol{Y} - \sum_{l=1}^{L} \beta_l \left [\boldsymbol{a}_{r} (\hat{r}_l) \odot \boldsymbol{a}_{v} (\hat{v}_l) \odot \boldsymbol{a}_{t} (\hat{\theta}_l) \right ] \boldsymbol{b}_{r} (\hat{\theta}_l)^T  \bigg\|_F^2.
\end{equation}
We solve the problem through traversing all possible combinations of the estimated target velocities within the set $\boldsymbol{\hat{v}}$, the estimated target DOAs within the set $\boldsymbol{\hat{\theta}}$ and the estimated target ranges within the set $\boldsymbol{\hat{r}}$. The received signal is projected onto the signal space constructed by each 3D-parameter combination, and the parameters of the estimated targets are determined as the combination corresponding to the minimum residual. 

\section{CRLB of The Target Parameter Estimation} \label{sec4}

For the model (\ref{model}), we derive the CRLB for the estimation of the target range, DOA and velocity. We first transfer the model into vector from as
\begin{equation}
	\boldsymbol{\tilde{y}} = \sum_{l=1}^{L} \beta_l \boldsymbol{b}_r (\theta_l) \otimes \left [\boldsymbol{a}_r (r_l) \odot \boldsymbol{a}_{v} (v_l) \odot \boldsymbol{a}_{t} (\theta_l) \right ].
\end{equation}
Assume that $\sigma_n^2$ denotes the variance of Gaussian white noise added on $\boldsymbol{\tilde{y}}$. $\boldsymbol{\tilde{y}}$ then obeys Gaussian distribution, i.e.,
\begin{equation}
	\boldsymbol{\tilde{y}} \sim \mathcal 
	{CN} ( \sum_{l=1}^{L} \beta_l \boldsymbol{b}_r (\theta_l) \otimes \left [\boldsymbol{a}_r (r_l) \odot \boldsymbol{a}_{v} (v_l) \odot \boldsymbol{a}_{t} (\theta_l) \right ], \boldsymbol{C} ),
\end{equation}
where $\boldsymbol{C} = \sigma_n^2 \boldsymbol{I}$, and $\boldsymbol{I} $ denotes an identity matrix. 

The parameters to be estimated are
\begin{equation}
	\boldsymbol{s} = \left[ \boldsymbol{\theta}^\text{T} \quad \boldsymbol{r}^\text{T} \quad \boldsymbol{v}^\text{T}  \right]^\text{T},
\end{equation}
where $\boldsymbol{\theta} \triangleq \left[\theta_1 \; \theta_2 \; \cdots \; \theta_L\right]^\text{T}$, $\boldsymbol{r} \triangleq \left[r_1 \; r_2 \; \cdots r_L \; \right]^\text{T}$, $\boldsymbol{v} \triangleq \left[v_1 \; v_2 \; \cdots \; v_L\right]^\text{T}$. The fisher matrix with respect to $\boldsymbol{s}$ is

%	\begin{equation}
	%		\begin{split}
		%			& \boldsymbol{F}(s_i,s_j)= -\mathbb{E} \{ \frac{\partial^2 \mathrm{ln} f(\boldsymbol{y} ; \boldsymbol{s})}{\partial s_i \partial s_j}\} .
		%		\end{split}
	%	\end{equation}
%% Assume that $\sigma_n^2$ denotes the variance of Gaussian white noise added on $\boldsymbol{\tilde{y}}$, the Fisher matrix can be further represented as

\begin{equation}
	\begin{split}
		\boldsymbol{F}(s_i,s_j) & = -\mathbb{E} \{ \frac{\partial^2 \mathrm{ln} f(\boldsymbol{y} ; \boldsymbol{s})}{\partial s_i \partial s_j}\}
		\\ & = \frac{2}{\sigma_n^2} \mathcal{R} \{ \mathrm{tr} [ \frac{\partial \boldsymbol{\tilde{y}}}{\partial s_i} \frac{\partial \boldsymbol{\tilde{y}}^\text{H}}{\partial s_j} ] \}
		\\ & =  \frac{2}{\sigma_n^2} \mathcal{R} \{ \frac{\partial \boldsymbol{\tilde{y}}^\text{H}}{\partial s_j} \frac{\partial \boldsymbol{\tilde{y}}}{\partial s_i}  \}
	\end{split}
\end{equation}
We then take a derivative with respect to $\theta_l$, $r_l$ and $v_l$, respectively.

\begin{equation}
	\begin{split}
		\frac{\partial \boldsymbol{\tilde{y}}}{\partial \theta_l} = & \beta_l \frac{\partial \boldsymbol{b}_r (\theta_l)}{\partial \theta_l} \otimes \left [\boldsymbol{a}_r (r_l) \odot \boldsymbol{a}_{v} (v_l) \odot \boldsymbol{a}_{t} (\theta_l) \right ]
		\\ & + \beta_l  \boldsymbol{b}_r (\theta_l) \otimes \left [\boldsymbol{a}_r (r_l) \odot \boldsymbol{a}_{v} (v_l) \odot \frac{\boldsymbol{a}_{t} (\theta_l)}{\theta_l} \right ]
	\end{split}
\end{equation}
\begin{equation}
	\frac{\partial \boldsymbol{\tilde{y}}}{\partial r_l} = \beta_l  \boldsymbol{b}_r (\theta_l) \otimes \left [\frac{\boldsymbol{a}_r (r_l)}{r_l} \odot \boldsymbol{a}_{v} (v_l) \odot \boldsymbol{a}_{t} (\theta_l) \right ]
\end{equation}
\begin{equation}
	\frac{\partial \boldsymbol{\tilde{y}}}{\partial v_l} = \beta_l  \boldsymbol{b}_r (\theta_l) \otimes \left [\boldsymbol{a}_r (r_l) \odot \frac{\boldsymbol{a}_v (v_l)}{v_l} \odot \boldsymbol{a}_{t} (\theta_l) \right ]
\end{equation}
Finally, the CRLB with respect to $\boldsymbol{s}_i$ is 
\begin{equation} 
	\mathrm{CRLB} \{\boldsymbol{s}_i \} = \left[ \boldsymbol{F}^{-1} \right]_{i,i} , 
\end{equation} 
where $ \boldsymbol{F}^{-1} $ denotes the inverse of $\boldsymbol{F}$ and $ \left[ \cdot \right ] _{i,i}$ denotes the element at the $i$-th column and the $i$-th row of the matrix.

\section{Complexity Analysis} \label{sec5}

The computational complexity of the proposed algorithms are shown in Table~\ref{Complexity}.

\begin{table}[H] \caption{Computational Complexity}
	\begin{tabular}{|c|c|} 
		\hline
		\textbf{Algorithms}                        & \textbf{Computational complexity}                                     \\ \hline
		2-fold DANM           & \makecell[c]{$ \big( (NMPQ_r + NM + PQ_r)^2 (NM+PQ_r)^2 +$ \\ $ (NMPQ_r + NM + PQ_r) (NM+PQ_r)^3\big) $ \\ $ \times \max(PQ_r,MN)^{0.5}$ } \\ \hline
		DDANM       & \makecell[c]{$ \big( M(NPQ_r + N + PQ_r)^2 (N+PQ_r)^2 +$ \\ $ M(NPQ_r + N + PQ_r) (N+PQ_r)^3\big) $ \\ $\times \max(PQ_r,N)^{0.5}$ }            \\ \hline
		\makecell[c]{Iterative \\ optimization} & $((NMPQ_r)^3+NMPQ_rN_g)N_{i}    $                                      \\ \hline
		OMP                               & $NKQ_rN_g^3$                           \\ \hline                          
	\end{tabular}
	\label{Complexity}
\end{table}
$N_g$ represents the number of grid in each dictionary matrix concerning the target DOA, the target range and the target velocity. $N_i$ represents the number of iterations in the optimization using $\ell_1$ norm minimization. The practical computational time of the algorithms are shown in Table~\ref{time}. All the simulation results are obtained on a PC with Matlab R2021b with a 3.6 GHz Intel Core i7.

% Assume that the target range $r_l \in [0,60m]$, the target DOA $\theta_l \in [-30^\circ,30^\circ]$ and the target velocity $v_l \in [-30m/s,30m/s]$.
\begin{table}[h] \caption{Computational Time}
	\centering
	\begin{tabular}{|c|cll|}
		\hline
		\multicolumn{1}{|l|}{} & \multicolumn{3}{l|}{\textbf{Computational Time / s}} \\ \hline
		2-fold DANM             & \multicolumn{3}{c|}{9.81}                   \\ \hline
		DDANM         & \multicolumn{3}{c|}{4.67}                   \\ \hline
		$\ell_1$ norm minimization  & \multicolumn{3}{c|}{3.80}                   \\ \hline
		OMP                    & \multicolumn{3}{c|}{965}                    \\ \hline
	\end{tabular}
	\label{time}
\end{table}

When solving the SDP problem with interior point method, the semi-definite constraints are transferred into the minimization of determinants. Newton method is then utilized to determine the extreme point in each iteration, where Hessian matrix inverse is necessary. Through applying DDANM algorithm, the determinant's dimensions are significantly reduced, leading to a notable reduction in computational complexity. As shown in~\cref{Complexity} and~\cref{time}, effective reductions in both the computational complexity and the computation time are observed.

%	We need to take the second derivatives of the determinant to all relevant variables, where the matrix inverse is required. 

%As described above, decomposed ANM splits the 3Dimensional matrix into multiple 2-dimensional matrix, and the computational complexity of matrix inverse is greatly reduced, which is

%in accordance with the comparison of decomposed ANM and 2-fold DANM in Table.~\ref{Complexity} and Table.~\ref{time}. 
$\ell_1$ norm minimization algorithm has the minimum computation time in~\cref{time}, while its theoretical computational complexity is not that low. The difference can potentially be attributed to the sparsity in the selection matrix. With few non-zero entries in the selection matrix, the computational complexity of solving the QP problem is much lower than expectation. 

%	 When simulating the computational time of OMP algorithm, we set $N_g = 1000$. 
The computational time of OMP algorithm is considerably higher than other algorithms with $N_g$ set as $1000$. If we decrease the number of grids, the estimation result of OMP will suffer from an error floor caused by the grid mismatch. The trade-off between computation complexity and estimation performance is inevitable in OMP algorithm.

\section{Simulation Results} \label{sec6}

\begin{table}[h]
	\centering
	\renewcommand{\arraystretch}{1.2}
	\caption{Simulation Parameters.}\label{table1}
	\begin{tabular}{cc}
		\hline
		\multicolumn{1}{c}
		{\textbf{Parameter}}                       & \multicolumn{1}{c}{Value} \\ \hline
		The number of transmitting antennas $P$                          & $8$                          \\
		The number of receiving antennas $Q_r$                             & $2$                          \\
		The number of targets S                                        & $1$                          \\
		The space between antennas d                                   & $0.5$ wavelengths             \\
		The number of pulses $N$                        & $20$                       \\
		The number of active antennas $K$ & $4$ \\
		The number of possible frequencies $M$ & $4$ \\
		The carrier frequency $f_0$                                     & $77$~GHz                    \\
		The default frequency difference $ \Delta f$               & $2.5$~MHz                    \\
		The default range of target $ r$               & $15,30,45$~m                    \\
		The default DOA of target $ \theta$               & $-30,10,40$~°                    \\
		The default velocity of target $v$   & $10,-20,20$~m/s \\
		The default SNR & 20dB \\
		\hline 
	\end{tabular}
\end{table}

%\begin{figure}[h] 
%	\centering
%	\includegraphics[width=0.475\textwidth]{D3.eps}
%	\caption{Range, DOA and velocity estimation result.}
%	\label{D3}
%\end{figure}

Simulation parameters are given in Table~\ref{table1}. The number of Monte Carlo simulations is $ 10^3$. The involved algorithm in the simulation includes:
\begin{itemize}
        \setlength{\parsep}{-10pt}
	\item 2-fold DANM \& HOOL:  Solve the problem formulated with 2-fold DANM algorithm (\ref{2-fold}) first, and use HOOI to decompose the optimized tensor. The algorithm is given in \cref{algorithm1}.
	\item 2-fold DANM \& Matching:  Solve the problem formulated with 2-fold DANM algorithm (\ref{2-fold}) first, and the target parameters are estimated with the optimized Toeplitz matrices. The target parameters matching is completed through traversing and projecting (\ref{prj1}). The algorithm is given in \cref{algorithm2}.
	\item DDANM \& HOOL:  Solve the problem formulated with DDANM algorithm (\ref{opt}) first, and use HOOI to decompose the optimized tensor.
	\item DDANM \& Matching:  Solve the problem formulated with DDANM algorithm (\ref{opt}) first, and the target parameters are estimated with the optimized Toeplitz matrices. The target parameters matching is completed through traversing and projecting (\ref{prj1}).
	\item $\ell_1$ norm minimization: Solve the problem formulated with $\ell_1$ norm minimization (\ref{l1}). The target parameters matching is completed through traversing and projecting (\ref{prj2}).
	\item OMP: OMP algorithm is used for target parameters estimation in FRaC system in \cite{frac}.
\end{itemize}
\begin{figure}[h] 
	\centering
	\includegraphics[width=0.425\textwidth]{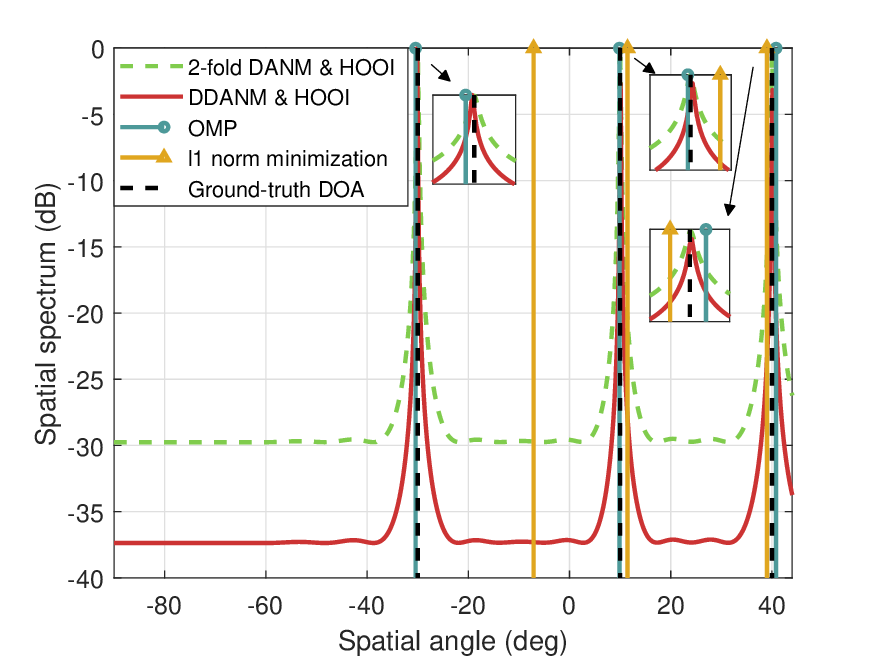}
	\caption{The spatial spectrum for target DOA estimation.}
	\label{sp}
\end{figure}

We first use spatial spectrum as an example to show the estimation performance of the proposed algorithms and the traditional OMP algorithm. As shown in Fig.~\ref{sp}, the performance of 2-fold DANM \& HOOL, DDANM \& HOOI and OMP algorithm are similar with the estimation results close to the ground-truth DOA. They all outperform $\ell_1$ norm minimization algorithm, which does not guarantee the detection of all the 3 targets. 

\begin{figure}[h] 
	\centering
	\includegraphics[width=0.425\textwidth]{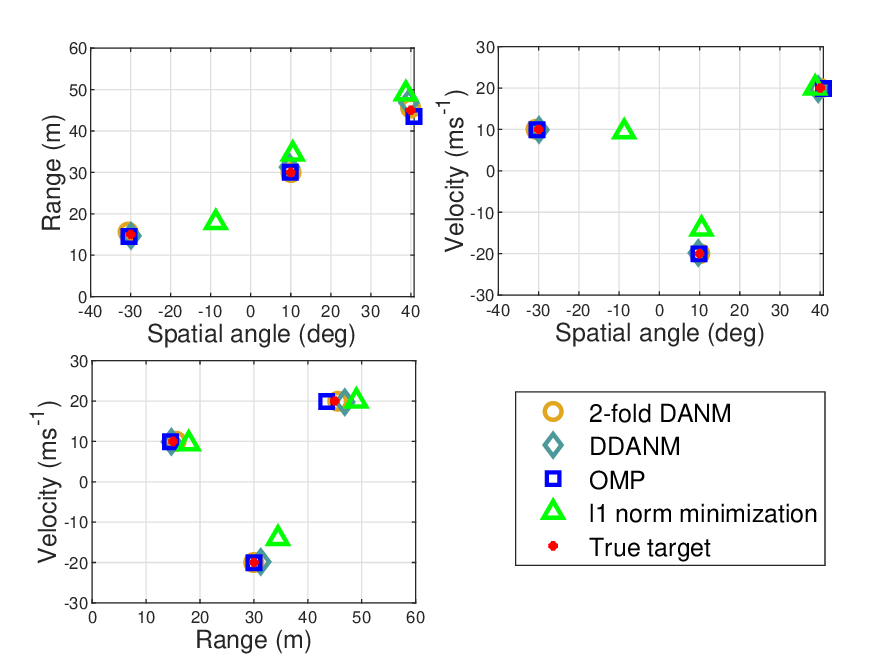}
	\caption{Range, DOA and velocity estimation results.}
	\label{D3}
\end{figure}

We then show the range, DOA and velocity estimation result of different algorithms in Fig.~\ref{D3}. The estimated range, DOA and velocity of 2-fold DANM \& HOOL, decomposed ANM \& HOOI and OMP algorithm are all close to the true range, DOA and velocity of the target. Moreover, the estimated parameters are accurately matched for the targets, which proves the effectiveness of the parameter matching and the estimation accuracy of the proposed algorithms. Besides, the result also verifies that the virtual aperture is fully utilized since the $3$ targets are successfully estimated with only $2$ receiving antenna at the receiver.

\begin{figure}[H] 
        \setlength{\textfloatsep}{-60pt}
	\subfloat[Range estimation]{
		\label{R_SNR}
		\includegraphics[width=0.425\textwidth]{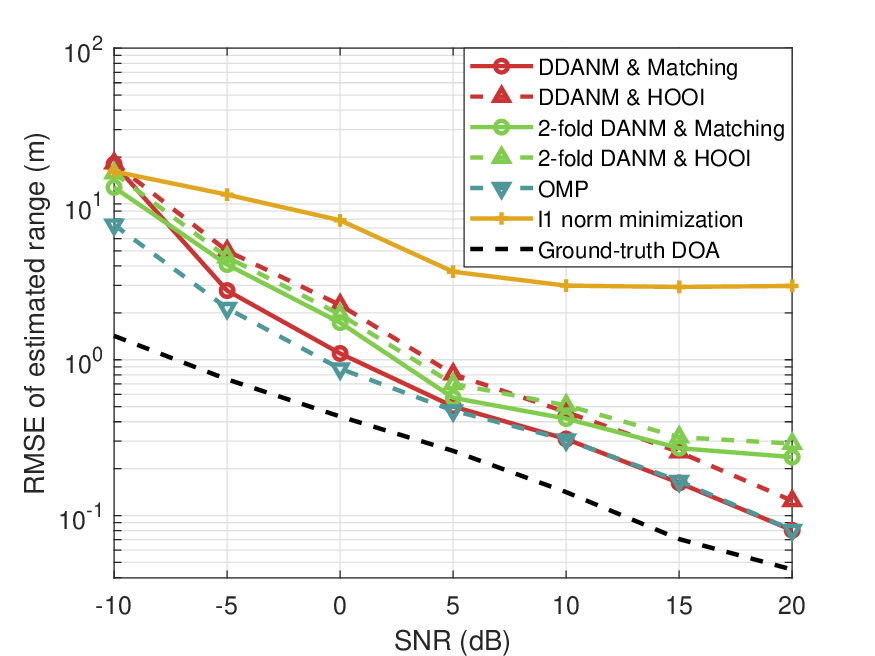}}
	\quad
	\subfloat[DOA estimation]{
		\label{DOA_SNR}  
		\includegraphics[width=0.425\textwidth]{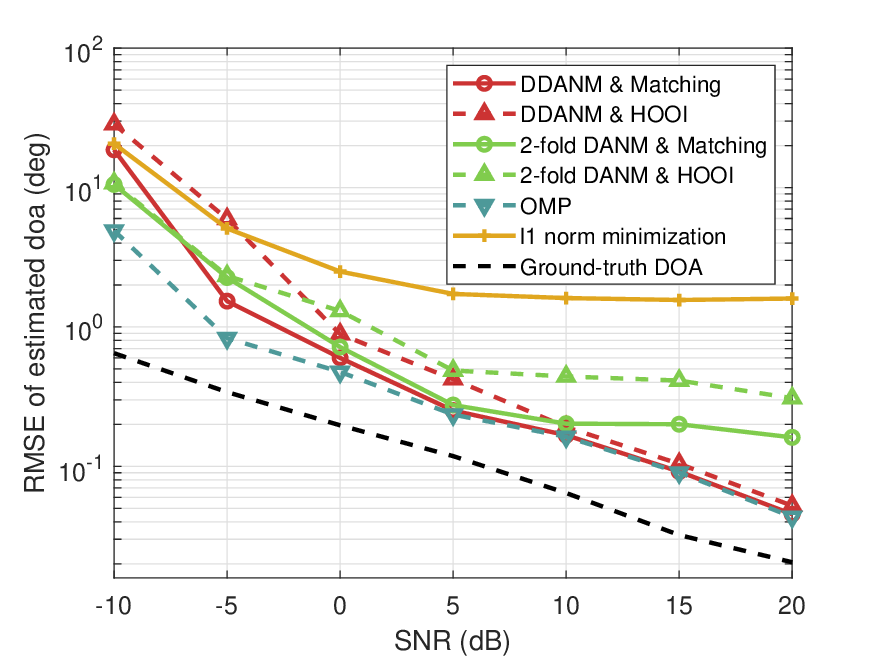}}
	\quad
	\subfloat[Velocity estimation]{
		\label{V_SNR}  
		\includegraphics[width=0.425\textwidth]{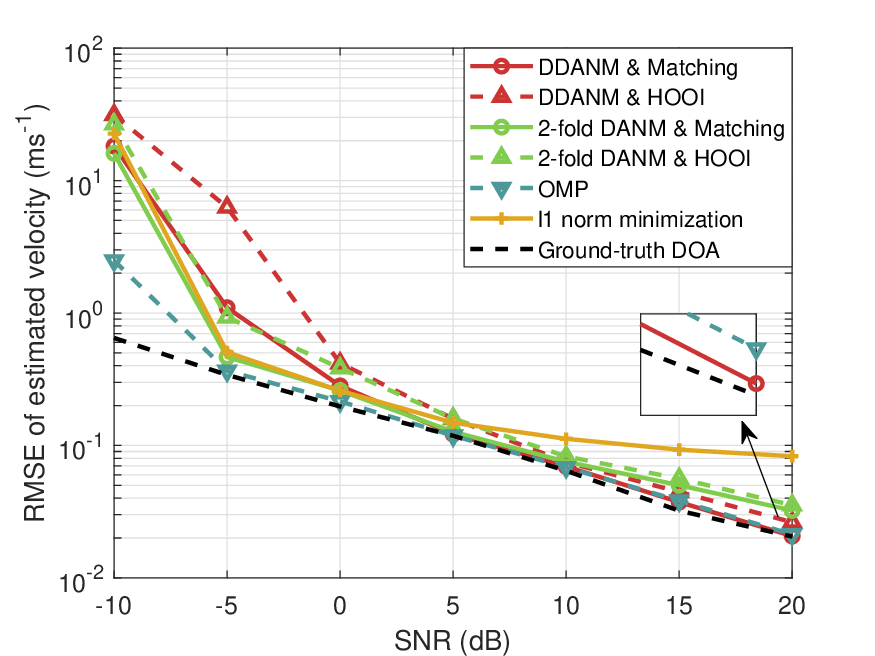}}
	\caption{RMSE versus SNR for range-angle-velocity estimation in single target scenario.}
	\label{fig_SNR}
\end{figure}

%

%\begin{figure}[htbp] 
%	\subfloat[Range estimation]{
%		\label{R_SNR}
%		\includegraphics[width=0.225\textwidth]{R_DOA.eps}}
%	\quad
%	\subfloat[DOA estimation]{
%		\label{DOA_SNR}  
%		\includegraphics[width=0.225\textwidth]{V_DOA.eps}}
%	\quad
%	\subfloat[Velocity estimation]{
%		\label{V_SNR}  
%		\includegraphics[width=0.225\textwidth]{R_V.eps}}
%	\caption{RMSE versus SNR for range-angle-velocity estimation in single target scenario.}
%	\label{fig_SNR}
%\end{figure}
Then, we show the relation between SNR and RMSE of the estimated parameters in single target scenario in Fig.~\ref{fig_SNR}\subref{R_SNR}, Fig.~\ref{fig_SNR}\subref{DOA_SNR} and Fig.~\ref{fig_SNR}\subref{V_SNR}. OMP algorithm and DDANM \& Matching algorithm performs the best. The performance of DDANM \& Matching algorithm is basically the same as OMP algorithm with high SNR but is seriously affected by noise. When SNR is less than 5dB, the performance of DDANM \& Matching algorithm decreases fast. DDANM \& HOOI algorithm performs wore than DDANM \& Matching algorithm. Tucker decomposition with HOOI algorithm introduces extra errors when estimating the target parameters with the optimized decoupled tensor compared with using the Toeplitz matrices as in DDANM \& Matching algorithm. 2-fold DANM algorithm performs better than DDANM algorithm with high SNR but has an error floor in both range and DOA estimation, while DDANM algorithm does not suffer from the error floor problem. Although $\ell_1$ norm minimization algorithm has the least computational complexity, it performs the worst when estimating the target range and DOA. The poor performance of $\ell_1$ norm minimization algorithm can be attributed to the non-ideal characteristic of the selection matrix for compressed sensing, and the reconstruction of the signal matrix can suffer from it. 

\begin{figure}[h] 
	\centering
	\includegraphics[width=0.425\textwidth]{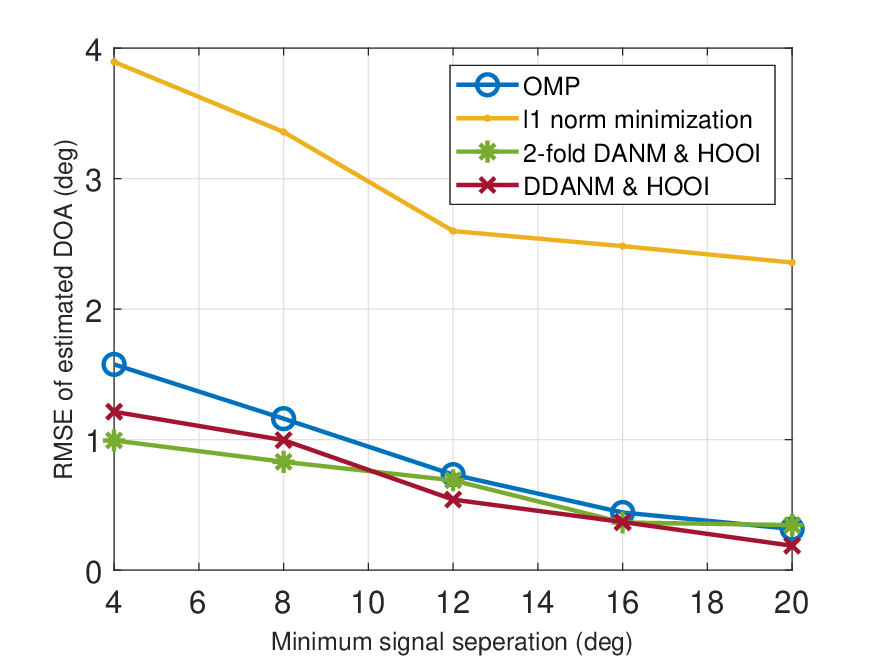}
	\caption{The DOA estimation with different minimum signal separations.}
	\label{sep}
\end{figure}

We then set the number of targets as 2 and show the DOA estimation performance with different minimum separations in Fig.~\ref{sep}. {Moreover, we give the angular resolution of FRaC system as a reference. According to the analysis in} \cite{frac}{, the FRaC system achieves the same range, velocity and angle resolution as that of a wideband FMCW radar system, which satisfy }
\begin{equation}
    \begin{aligned}
        \Delta r = \frac{c}{2M\Delta f}, \; \Delta v = \frac{\lambda}{2NT_0}, \; \Delta \theta &= \frac{2}{PQ_r \cos{\theta}}.
    \end{aligned}
\end{equation}
{With the simulation parameters given in Table.} \ref{table1}{, the angular resolution approximates ${14.3}^\circ$. According to Fig.}~\ref{sep}{, since the RMSE of the estimated DOA is far less than the signal separation even within the angular resolution, the used algorithms are proved to achieve super-resolution estimation.} 
The correlation between signals from different DOA is stronger when the DOA separation is smaller, and the estimation performance will be degraded. As shown in Fig.~\ref{sep}, the RMSE of the estimated DOA is far smaller than the DOA separation, which proves that  OMP, 2-fold DANM and DDANM algorithm are capable of reaching super-resolution estimation.

\begin{figure}[H]
	\subfloat[Range estimation]{
		\label{R_SNR2}
		\includegraphics[width=0.425\textwidth]{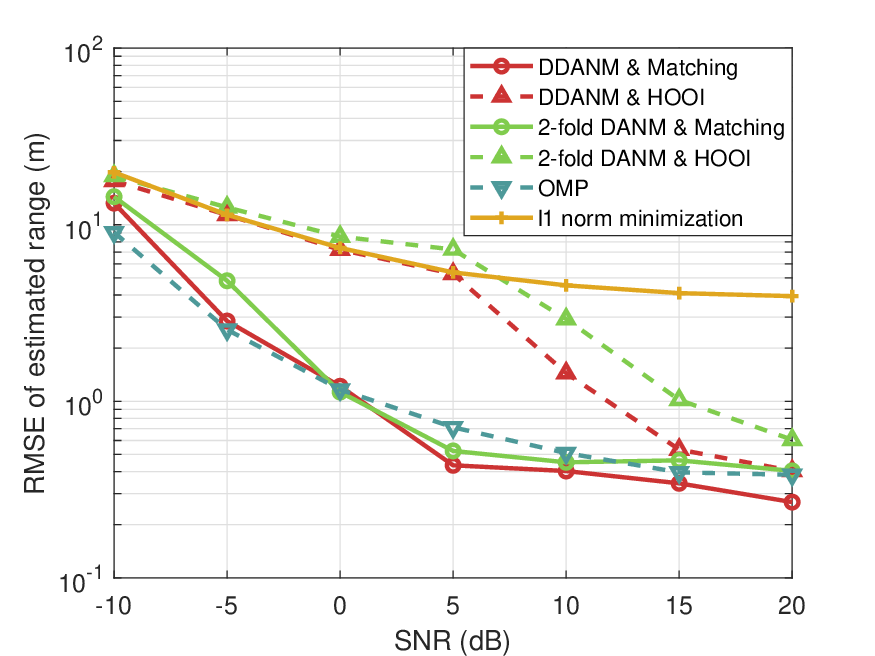}}
	\quad
	\subfloat[DOA estimation]{
		\label{DOA_SNR2}  
		\includegraphics[width=0.425\textwidth]{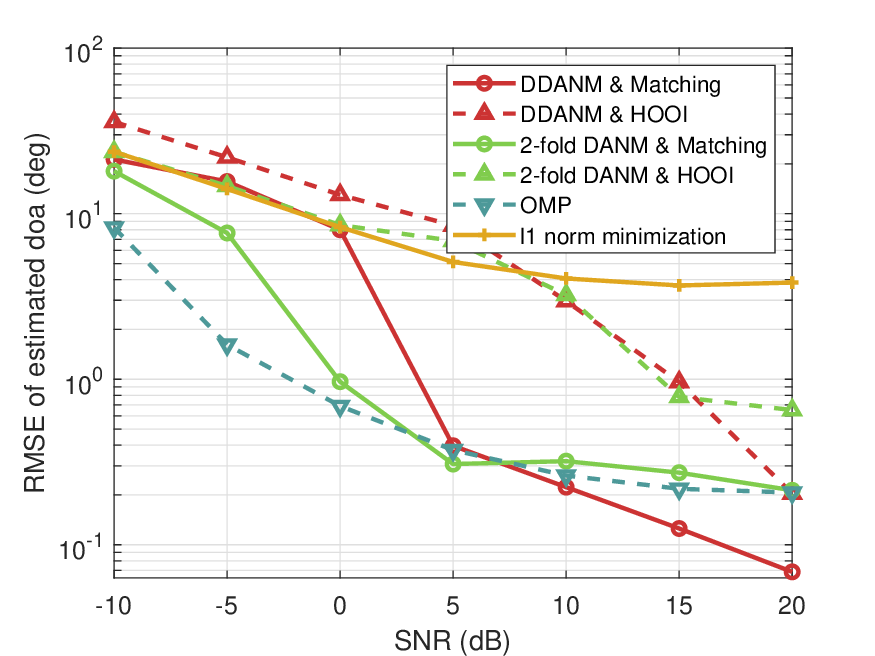}}
	\quad
	\subfloat[Velocity estimation]{
		\label{V_SNR2}  
		\includegraphics[width=0.425\textwidth]{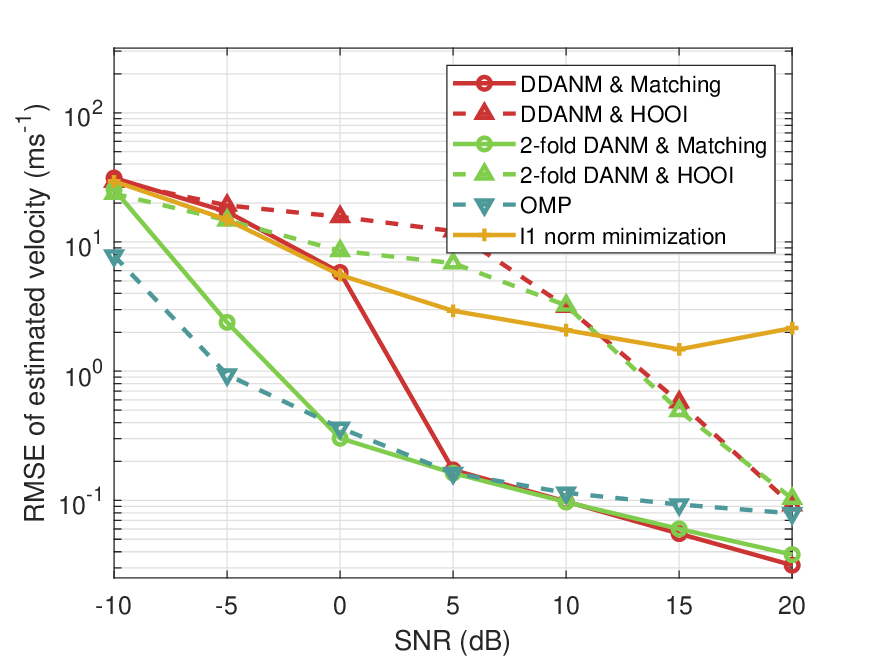}}
	\caption{RMSE versus SNR for range-angle-velocity estimation with 2 targets.}
	\label{RMSE2}
\end{figure}

%DDANM algorithm splits the $3$-dimensional matrix into multiple $2$-dimensional matrix and enable each splitted matrix to satisfy semi-definite constraint.

The estimation performance in 2 targets scenario is shown in Fig.~\ref{RMSE2}. DDANM \& Matching algorithm demonstrates superior performance when SNR exceeds $\text{10dB}$, and 2-fold DANM \& matching algorithm exhibits competitive performance within the range of $\text{10dB} > \text{SNR} >\text{0dB}$. When SNR is lower than 0dB, OMP algorithm has the minimum RMSE of estimation. Besides, the estimation performance of $\ell_1$ norm minimization algorithm is further degraded. The performance of the algorithms using HOOI algorithm also deteriorates. Even with the accurate optimized signal apace, the steering vector allocated for each target can still have errors with HOOI algorithm, which shows that Tucker decomposition does not align well with the proposed decoupling scheme. {The errors can arise from the non-uniqueness of Tucker decomposition. Unlike the PARAFAC decomposition, where uniqueness can be ensured under specific conditions related to the Kruskal rank of factor matrices, Tucker decomposition lacks a similar straightforward constraint for guaranteeing uniqueness} \cite{PARAFAC_unique,Tucker_unique}{. The ambiguity in decomposition leads to discrepancies in the derived factor matrices, diverging from the desired steering vectors. As a result, although the factor matrices can have the same vector space as the steering vectors, the vectors in the decomposed factor matrices cannot correspond one-to-one with the steering vectors.}

\section{Conclusion} \label{sec7}

The target range-DOA-velocity estimation in FRaC system have been considered in this paper. A decoupling scheme has been proposed to mitigate the coupling introduced by sparse MIMO array and IM technique, and the DDANM algorithm has been proposed as a computationally efficient algorithm. CRLB of the range-DOA-velocity estimation have been derived. The computational complexity has been analyzed and the computation time has been presented. Simulation results prove the effectiveness of the proposed scheme, which successfully decouples the coupled parameters while simultaneously harnessing the benefits afforded by virtual aperture exploitation. The proposed DDANM algorithm outperforms OMP algorithm and 2-fold DANM algorithm at a high SNR, which indicates that the proposed DDANM algorithm holds an advantage in terms of both complexity and estimation accuracy.

\bibliographystyle{IEEEtran}
%	\biboptions{sort&compress}
\bibliography{ref}
\clearpage
\begin{IEEEbiography}
[{\includegraphics[width=1in,height=1.25in,clip,keepaspectratio]{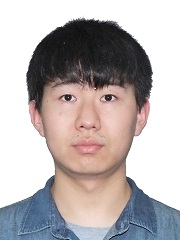}}] 
{Mengjiang Sun (Student Member, IEEE)} was born in Inner Mongolia, China, in 1998. He received the B.E. degree from School of Information Science and Engineering, Southeast University, China in 2021. He is currently pursing the Ph.D degree with the State Key Laboratory of Millimeter Waves, Southeast University, Nanjing, China.
His research interests include radar signal processing and millimeter wave communication.
\end{IEEEbiography}
\begin{IEEEbiography}
[{\includegraphics[width=1in,height=1.25in,clip,keepaspectratio]{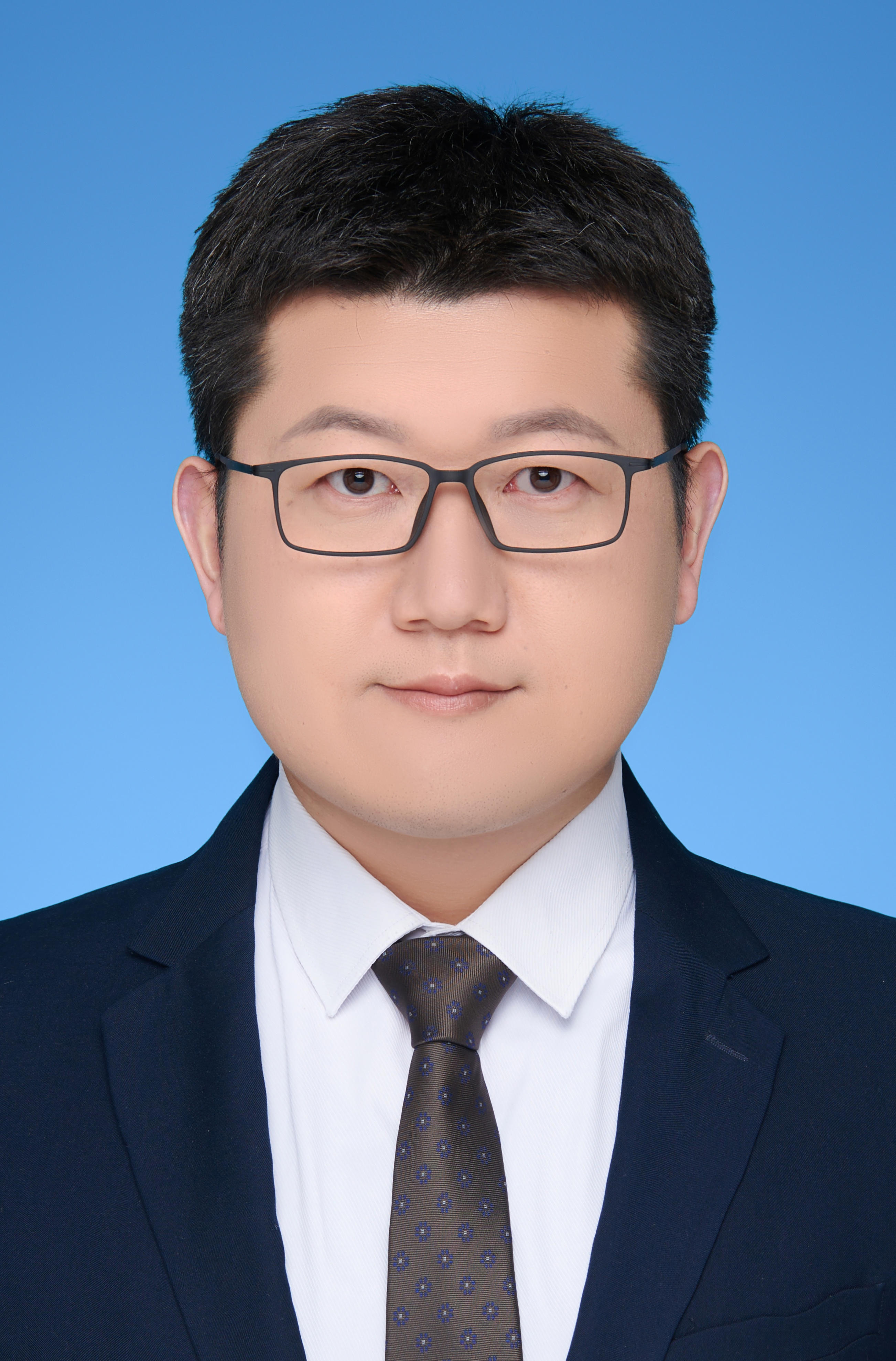}}] 
{Peng Chen (Seinor Member, IEEE)} received the B.E. and Ph.D. degrees from the School of Information Science and Engineering, Southeast University, Nanjing, China, in 2011 and 2017 respectively. From March 2015 to April 2016, he was a Visiting Scholar with the Department of Electrical Engineering, Columbia University, New York, NY, USA. He is currently an Associate Professor with the State Key Laboratory of Millimeter Waves, Southeast University. His research interests include target localization, super-resolution reconstruction, and array signal processing. He is a Jiangsu Province Outstanding Young Scientist. He has served as an IEEE ICCC Session Chair, and won the Best Presentation Award in 2022 (IEEE ICCC). He was invited as a keynote speaker at the IEEE ICET in 2022. He was recognized as an exemplary reviewer for IEEE WCL in 2021, and won the Best Paper Award at IEEE ICCCCEE in 2017. 
\end{IEEEbiography}
\begin{IEEEbiography}
[{\includegraphics[width=1in,height=1.25in,clip,keepaspectratio]{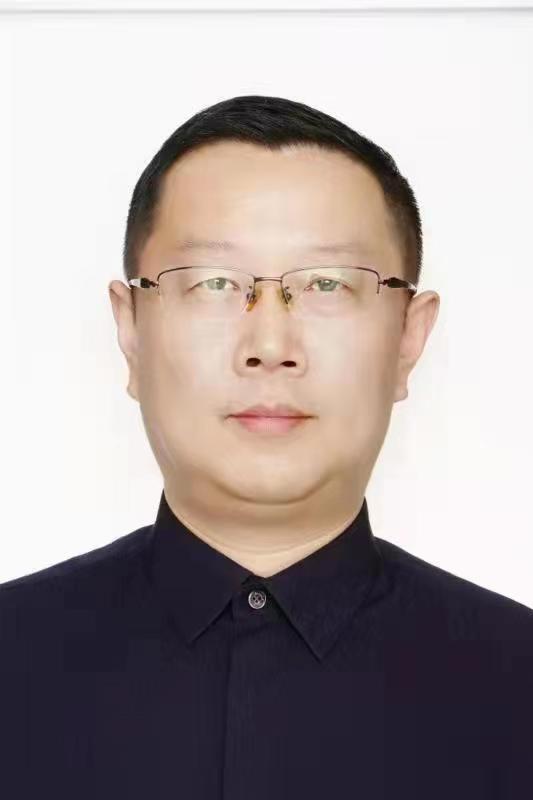}}] 
{Zhenxin Cao (Member, IEEE)} was born in May 1976. He received the M.S. degree from Nanjing University of Aeronautics and Astronautics, Nanjing, China, in 2002 and the Ph.D. degree from the School of Information Science and Engineering, Southeast University, Nanjing, China, in 2005. From 2012 to 2013, he was a Visiting Scholar with North Carolina State University. Since 2005, he has been with the State Key Laboratory of Millimeter Waves, Southeast University, where he is currently a Professor. His research interests include antenna theory and application.
\end{IEEEbiography}
\begin{IEEEbiography}
[{\includegraphics[width=1in,height=1.25in,clip,keepaspectratio]{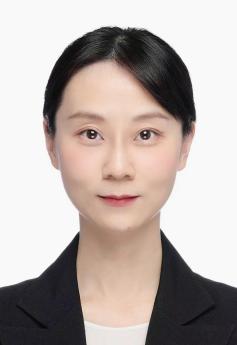}}] 
{Fei Shen (Member, IEEE)} is a Professor with Shanghai Institute of Microsystem and Information Technology (SIMIT), Chinese Academy of Sciences (CAS). She received her B.Eng. degree in Information Technology from Southeast University, China, and Ph.D. degree from Dresden University of Technology (TU Dresden), Germany. She worked as a  scientific research fellow in the Chinese University of Hong Kong (CUHK) , Ulm University and TU Dresden, accomplishing the Priority Program “COIN” funded by German Research Foundation (DFG). During 2015 to 2017, she worked as a Postdoc researcher in CentraleSupelec, Paris-Saclay University, and then a senior research engineer in Telecom, Institut Polytechnique de Paris, France. Prof. Shen is the Editor for IEEE Trans. on Network Science and Technology (TNSE) and Springer Journal on Wireless Personal Communications (WPS). She has been the PI for tens of national and provincial projects. Her current research interests include wireless communications, edge and fog computing, metaverse technologies.
\end{IEEEbiography}

\end{document}